
\input harvmac.tex
\overfullrule=0mm
\hfuzz 10pt
\input epsf.tex
\newcount\figno
\figno=0
\def\fig#1#2#3{
\par\begingroup\parindent=0pt\leftskip=1cm\rightskip=1cm\parindent=0pt
\baselineskip=11pt
\global\advance\figno by 1
\midinsert
\epsfxsize=#3
\centerline{\epsfbox{#2}}
\vskip 10pt
{\bf Fig. \the\figno:} #1\par
\endinsert\endgroup\par
}
\def\figlabel#1{\xdef#1{\the\figno}}
\def\encadremath#1{\vbox{\hrule\hbox{\vrule\kern8pt\vbox{\kern8pt
\hbox{$\displaystyle #1$}\kern8pt}
\kern8pt\vrule}\hrule}}
%
%
%
%
\def\frac#1#2{{\scriptstyle{#1 \over #2}}}

%
%

%
\def\({ \left( }\def\[{ \left[ }
\def\){ \right) }\def\]{ \right] }
%


\def\IR{\relax{\rm I\kern-.18em R}}
\font\cmss=cmss10 \font\cmsss=cmss10 at 7pt
\def\IZ{\relax\ifmmode\mathchoice
{\hbox{\cmss Z\kern-.4em Z}}{\hbox{\cmss Z\kern-.4em Z}}
{\lower.9pt\hbox{\cmsss Z\kern-.4em Z}}
{\lower1.2pt\hbox{\cmsss Z\kern-.4em Z}}\else{\cmss Z\kern-.4em Z}\fi}
\def\inbar{\,\vrule height1.5ex width.4pt depth0pt}
\def\IB{\relax{\rm I\kern-.18em B}}
\def\IC{\relax\hbox{$\inbar\kern-.3em{\rm C}$}}
\def\ID{\relax{\rm I\kern-.18em D}}
\def\IE{\relax{\rm I\kern-.18em E}}
\def\IF{\relax{\rm I\kern-.18em F}}
\def\IG{\relax\hbox{$\inbar\kern-.3em{\rm G}$}}
\def\IH{\relax{\rm I\kern-.18em H}}
\def\II{\relax{\rm I\kern-.18em I}}
\def\IK{\relax{\rm I\kern-.18em K}}
\def\IL{\relax{\rm I\kern-.18em L}}
\def\IM{\relax{\rm I\kern-.18em M}}
\def\IN{\relax{\rm I\kern-.18em N}}
\def\IO{\relax\hbox{$\inbar\kern-.3em{\rm O}$}}
\def\IP{\relax{\rm I\kern-.18em P}}
\def\IQ{\relax\hbox{$\inbar\kern-.3em{\rm Q}$}}
\def\IGa{\relax\hbox{${\rm I}\kern-.18em\Gamma$}}
\def\IPi{\relax\hbox{${\rm I}\kern-.18em\Pi$}}
\def\ITh{\relax\hbox{$\inbar\kern-.3em\Theta$}}
\def\IOm{\relax\hbox{$\inbar\kern-3.00pt\Omega$}}


\def\oh{{1\over 2}}

\def\Ga{\alpha}\def\Gb{\beta}


\def\bra{\langle}\def\ket{\rangle}

\def\\#1 {{\tt\char'134#1} }

\catcode`\@=11
\def\Eqalign#1{\null\,\vcenter{\openup\jot\m@th\ialign{
\strut\hfil$\displaystyle{##}$&$\displaystyle{{}##}$\hfil
&&\qquad\strut\hfil$\displaystyle{##}$&$\displaystyle{{}##}$
\hfil\crcr#1\crcr}}\,}   \catcode`\@=12
\def\encadre#1{\vbox{\hrule\hbox{\vrule\kern8pt\vbox{\kern8pt#1\kern8pt}
\kern8pt\vrule}\hrule}}
\def\encadremath#1{\vbox{\hrule\hbox{\vrule\kern8pt\vbox{\kern8pt
\hbox{$\displaystyle #1$}\kern8pt}
\kern8pt\vrule}\hrule}}


%
\newdimen\xraise\newcount\nraise
\def\xpoint{\hbox{\vrule height .45pt width .45pt}}
\def\udiag#1{\vcenter{\hbox{\hskip.05pt\nraise=0\xraise=0pt
\loop\ifnum\nraise<#1\hskip-.05pt\raise\xraise\xpoint
\advance\nraise by 1\advance\xraise by .4pt\repeat}}}
\def\ddiag#1{\vcenter{\hbox{\hskip.05pt\nraise=0\xraise=0pt
\loop\ifnum\nraise<#1\hskip-.05pt\raise\xraise\xpoint
\advance\nraise by 1\advance\xraise by -.4pt\repeat}}}
\def\vertex{\epsfxsize=5mm\hbox{\raise -1mm\hbox{\epsfbox{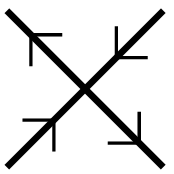}}}}
\def\vertexalt{\epsfxsize=5mm\hbox{\raise -1mm\hbox{\epsfbox{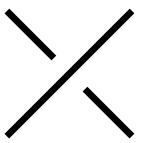}}}}
\def\vertexdbl{\epsfxsize=8mm\hbox{\raise -4mm\hbox{\epsfbox{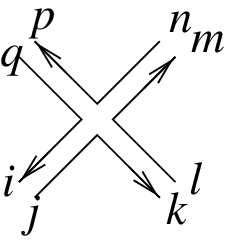}}}}
\def\propagdbl{\epsfxsize=12mm\hbox{\raise -1mm\hbox{\epsfbox{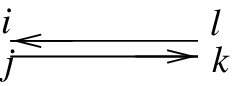}}}}
\def\vertexarr{\epsfxsize=8mm\hbox{\raise -4mm\hbox{\epsfbox{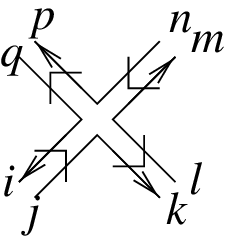}}}}
\def\propagarr{\epsfxsize=12mm\hbox{\raise -1mm\hbox{\epsfbox{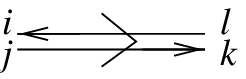}}}}
\def\Th{Thistlethwaite}
\def\c{{\,\rm c}}
\def\ommit#1{}
%
%
%
\lref\AV{I.Ya. Arefeva and I.V. Volovich, 
{\sl Knots and Matrix Models}, {\it Infinite Dim.
Anal. Quantum Prob.} 1 (1998) 1 ({\tt hep-th/9706146}).}
\lref\BIPZ{E. Br\'ezin, C. Itzykson, G. Parisi and J.-B. Zuber, 
{\sl Planar Diagrams}, {\it Commun. Math. Phys.} {\bf 59} (1978) 35--51.}
\lref\BIZ{D. Bessis, C. Itzykson and J.-B. Zuber, 
{\sl Quantum Field Theory Techniques in Graphical Enumeration},
{\it Adv. Appl. Math.} {\bf 1} (1980) 109--157.}
\lref\DFGZJ{P. Di Francesco, P. Ginsparg and J. Zinn-Justin, 
{\sl 2D Gravity and Random Matrices, }{\it Phys. Rep.} {\bf 254} (1995)
1--133.}
\lref\tH{G. 't Hooft, 
{\sl A Planar Diagram Theory for Strong 
Interactions}, {\it Nucl. Phys.} {\bf B 72} (1974) 461--473.}
\lref\HTW{J. Hoste, M. Thistlethwaite and J. Weeks, 
{\sl The First 1,701,936 Knots}, {\it The Mathematical Intelligencer}
{\bf 20} (1998) 33--48.}
\lref\MTh{W.W. Menasco and M.B. \Th, 
{\sl The Tait Flyping Conjecture}, {\it Bull. Amer. Math. Soc.} {\bf 25}
(1991) 403--412; 
{\sl The Classification of Alternating 
Links}, {\it Ann. Math.} {\bf 138} (1993) 113--171.}
\lref\Ro{D. Rolfsen, {\sl Knots and Links}, Publish or Perish, Berkeley 1976.}
\lref\STh{C. Sundberg and M. Thistlethwaite, 
{\sl The rate of Growth of the Number of Prime Alternating Links and 
Tangles}, {\it Pac. J. Math.} {\bf 182} (1998) 329--358.}
\lref\Tutte{W.T. Tutte, {\sl A Census of Planar Maps}, 
{\it Can. J. Math.} {\bf 15} (1963) 249--271.}
\lref\Zv{A. Zvonkin, {\sl Matrix Integrals and Map Enumeration: An Accessible
Introduction},
{\it Math. Comp. Modelling} {\bf 26} (1997) 281--304.}
\lref\KM{V.A.~Kazakov and A.A.~Migdal, {\sl Recent progress in the
theory of non-critical strings}, {\it Nucl. Phys.} {\bf B 311} (1988)
171--190.}
\lref\KP{V.A.~Kazakov and P.~Zinn-Justin, {\sl Two-Matrix Model with
$ABAB$ Interaction}, {\it Nucl. Phys.} {\bf B 546} (1999) 647
({\tt hep-th/9808043}).}
\Title{
\vbox{\baselineskip12pt\hbox{RU-99-16}\hbox{SPhT 99/037}\hbox{{\tt math-ph/9904019}}}}
{{\vbox {
\vskip-10mm
\centerline{Matrix Integrals and the Counting of Tangles and Links}
}}}
\medskip
\centerline{P. Zinn-Justin}\medskip
\centerline{\it Department of Physics and Astronomy,  Rutgers University,} 
\centerline{\it Piscataway, NJ 08854-8019, USA}
\bigskip
\centerline{and}
\medskip
\centerline{J.-B. Zuber}\medskip
\centerline{\it C.E.A.-Saclay, Service de Physique Th\'eorique,}
\centerline{\it F-91191 Gif sur Yvette Cedex, France}

\vskip .2in

\noindent 
Using matrix model techniques for
the counting of planar Feynman diagrams, recent results 
of Sundberg and \Th\ on the counting of alternating tangles
and links are reproduced.  

\bigskip
\vskip45mm

\noindent{to appear in the proceedings of the 11th 
International Conference on Formal Power Series and Algebraic 
Combinatorics, Barcelona June 1999}
\Date{4/99}
%

\newsec{Introduction }
\noindent
This is a paper of physical mathematics, which  means that it addresses a 
problem of mathematics using tools of (theoretical) physics.
The problem of mathematics is a venerable one, more than a hundred years old, 
namely the counting
of (topologically inequivalent) knots. The physical tools are 
combinatorial methods developed in the framework of field theory and
so-called matrix models. For a review of the history and recent 
developments of the first subject, see \HTW. For an introduction 
for non physicists to matrix integral techniques, see for example
\refs{\BIZ{--}\Zv}. 

In this note, we show that by combining results obtained recently 
in knot theory and older ones on matrix integrals, and by using 
graphical decompositions familiar in field theory,  one may 
reproduce and somewhat simplify the counting of alternating tangles and
links performed in \STh. In section 2, we recall 
basic facts and definitions on knots and their planar projections;
we also recall why integrals over large matrices are relevant 
for the counting of planar objects. 

Specifically, we shall consider the following integral
%
\eqn\Ia{\int dM dM^\dagger \exp-N\, \tr \left( \alpha  MM^\dagger 
-{g\over 2}(MM^\dagger)^2\right)\ .}
%
over $N\times N$ complex  matrices, in the large $N$ limit. 
In that limit, the integral is represented in terms of {\it planar} Feynman 
diagrams, with directed edges and four-valent vertices of the type 
\vertex, 
which exhibit a close similarity  with alternating knot 
diagrams in planar projection, with crossings represented as
\vertexalt.
Thus the counting of planar Feynman diagrams (with 
adequate conditions and weights) must be related to the counting of 
alternating knots. 
A substantial part of this paper (section 3) is devoted to eliminating 
irrelevant or redundant contributions of Feynman diagrams. Once
this is achieved, the results of \STh\ are recovered. 
In the concluding section, we comment on the possible extensions of
these methods.  

The observation that planar Feynman diagrams generated by 
matrix models can be associated to knot diagrams was
already made in \AV; the matrix integral proposed there
was more complicated,
so that no explicit calculation was carried out. 
 
%
\newsec{Basics} 
\subsec{Knots, links and tangles}
\noindent
In this section, we briefly recall some basic concepts of knot theory,
referring to the literature for more precise definitions.
A {\it knot} is a smooth circle embedded in $\IR^3$.
A {\it link} is a collection of intertwined knots
(in the following, we shall not consider ``unlinks'', i.e.\ links which can
be divided in several non-intertwined pieces). 
Both kinds of objects are considered up to homeomorphisms of $\IR^3$.
Roughly speaking, 
a {\it tangle} is a knotted structure from which four strings 
emerge: if this object is contained in a ball $B$ with the four endpoints 
of the strings attached on $\partial B$, 
topological equivalence is up to orientation preserving
homeomorphisms of $B$ that reduce to the identity on $\partial B$.  
The fundamental problem of knot theory is the classification 
of topologically inequivalent knots, links and tangles. 

\vskip -8pt
%
\fig{(a): a non prime link; (b): an irrelevant (or ``nugatory'') crossing;
(c): 2 particle-reducible tangles, horizontal or vertical sums of two tangles}
{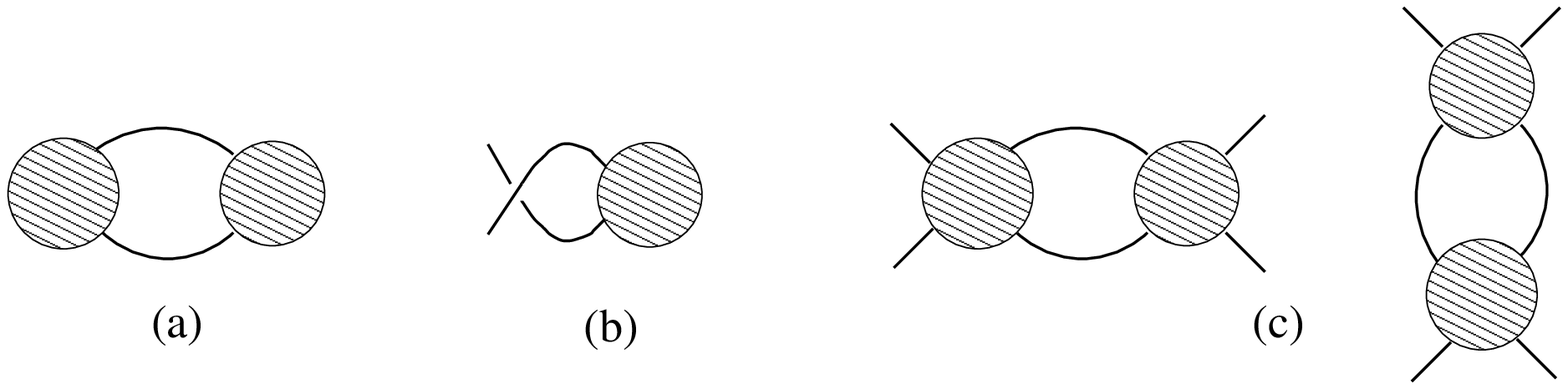}{7cm}\figlabel\nprime
It is common to represent such objects by their projection in a plane, 
with under/over-crossings at each double point and with the minimal number 
of such crossings. 
To avoid redundancies, we can concentrate on {\it prime}
links and tangles, whose diagrams cannot be
decomposed as a sum of components (Fig.~\nprime) and on 
{\it reduced} diagrams that contain no irrelevant crossing. 

A diagram is called {\it alternating} if one meets alternatively 
under- and over-crossings as one travels along each loop. 
Starting with eight (resp six) crossings, there are knot (resp link) 
diagrams that cannot be  drawn in an alternating way. Although 
alternate links (and tangles) constitute only a subclass (asymptotically
subdominant), they are easier to characterize and thus to enumerate or
to count. 
%
\fig{The flype of a tangle}
{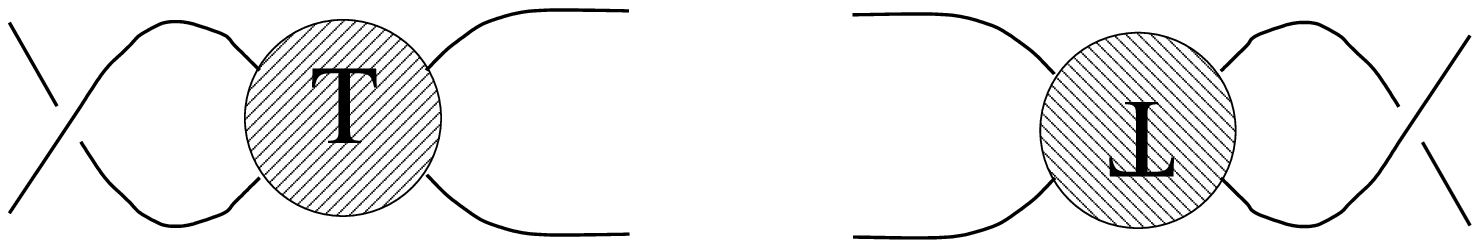}{5cm}\figlabel\flyp
\noindent 
A major result conjectured by Tait and proved  by Menasco and \Th~\MTh\
is that two alternating reduced knot or link diagrams on the sphere represent 
the same object if
and only if they are related by a sequence of moves acting on tangles 
called ``flypes'' (see Fig.~\flyp).
 
The subproblem that we shall address here is thus the counting of alternating 
prime links and tangles. 


\subsec{Matrix Integrals in the large $N$ limit}
\noindent 
As a prototype of matrix integrals, we 
consider the integral over $N\times N$ hermitian matrices
%
\eqn\IIa{ Z=\int dM \exp-N\, \tr \left(\oh  M^2
-{g\over 4} M^4\right)\ . } 
In order that the integral makes sense, the sign of $g$
should originally be chosen negative. As is well-known,
the large $N$ limit allows the analytic continuation
to positive values of $g$.
The series expansion of $Z$ in powers of $g$ may be represented 
diagrammatically by Feynman diagrams, made of undirected 
edges or ``propagators'' (2-point functions
of the Gaussian model) with double lines expressing the conservation
of indices,
$\bra M_{ij} M_{k\ell}\ket_0=\propagdbl={1\over N}
\delta_{i\ell}\delta_{jk}$, and of four-valent vertices
$\vertexdbl=gN \delta_{qi}\delta_{jk}
\delta_{\ell m}\delta_{np}$. In the large $N$ limit 
a counting of powers of $N$ shows that the leading contribution 
to $\log Z$ is given by a sum of 
diagrams that may be drawn on the sphere~\tH, called ``planar'' by 
abuse of language. More precisely
\eqn\IIb{\lim_{N\to \infty} {1\over N^2} \log Z
=\sum_{{\scriptstyle {\rm planar\ diagrams}}\atop 
{\scriptstyle {\rm with\ } n\ {\rm vertices}}} {\rm weight}\ \ g^n }
with a weight equal to one over the order of the automorphism group of
the diagram. Once this has been realised, it is simpler to return 
to a notation with simple lines ${{}\over \qquad}$ and rigid vertices
$\ddiag{30}\!\!\!\!\!\!\!\udiag{30}$.
It is this property of matrix integrals to generate
(weighted) sums over planar diagrams that we shall use in the context 
of knot theory. 

%
\newsec{From planar Feynman diagrams to links and tangles } 
\subsec{The matrix integral}
\noindent As mentionned in the Introduction, in the context of knot theory, 
it seems natural to consider the integral \Ia\ 
over complex (non hermitian) matrices, in order
to distinguish between under-crossings and over-crossings.
However, this integral is closely
related to the simpler integral \IIa\ in the large $N$ limit. 
Let us define the partition functions
\eqna\IIIa
$$\eqalignno{
Z^{(1)}(\alpha,g)&=\int dM dM^\dagger \exp-N\, \tr \left( \alpha  MM^\dagger 
-{g\over 2}(MM^\dagger)^2\right)&\IIIa a\cr
Z(\alpha,g)&=\int dM \exp-N\, \tr \left( \oh \alpha  M^2
-{g\over 4}M^4\right)&\IIIa b\cr
}$$
and the corresponding ``free energies''
\eqna\IIIaa
$$\eqalignno{
F^{(1)}(\alpha,g)&=\lim_{N\to\infty} {1\over N^2} {\log Z^{(1)}(\alpha,g)
\over\log Z^{(1)}(\alpha,0)}&\IIIaa a\cr
F(\alpha,g)&=\lim_{N\to\infty} {1\over N^2} {\log Z(\alpha,g)
\over\log Z(\alpha,0)}\ .&\IIIaa b\cr
}$$
The constant $\alpha$ can be absorbed in a rescaling $M\to \alpha^{-\oh} M$:
\eqna\IIIb
$$\eqalignno{
Z(\alpha,g)&=\alpha^{-{N^2\over 2}} Z\big(1,{g\over\alpha^2}\big)&\IIIb a\cr
F(\alpha, g)&= F\big(1,{g\over\alpha^2}\big)&\IIIb b\cr
}$$
and similarly for $Z^{(1)}$ and $F^{(1)}$.
However, it will be useful for our purposes to keep the parameter
$\alpha$.

The Feynman rules of \IIIa{a} and those of \IIIa{b} are quite similar: 
the former have already been described in Sect.~2.2, while the latter
are given by a propagator $\bra M_{ij} M^\dagger_{k\ell}\ket_0
=\propagarr ={1\over N\alpha}\delta_{i\ell}\delta_{jk}$ 
and a four-vertex \vertexarr\ equal to $gN$ times the usual delta functions. 
The only difference lies in the 
orientation of the propagator of the complex theory. 
However, in the large $N$ ``planar'' limit, 
this only results in
a factor of 2 in the corresponding free energies,
which accounts for the two possible 
overall orientations that may be given to the lines of each 
graph of the hermitian theory 
to transform it into a graph of the non-hermitian one. 
Therefore 
\eqn\IIIc{
F^{(1)}(\alpha,g)=2 F(\alpha,g) }

In addition to the partition functions and free energies, one is also
interested in the correlation functions $G_{2n}(\alpha,g)
=\bra {1\over N}\tr M^{2n}\ket$ and 
$G^{(1)}_{2n}(\alpha,g)=\bra {1\over N}\tr (MM^\dagger)^n\ket$. 
The first ones, namely 
the 2-point and 4-point functions, are simply expressed in terms of $F$.
\eqn\IIId%
{ G_4=4 {\partial F\over\partial g}  
\qquad \qquad G^{(1)}_4=2 {\partial F^{(1)}\over\partial g}
}
so that $G^{(1)}_4=G_4$ in the large $N$ limit. Furthermore,
combining \IIId{} with the homogeneity property \IIIb{}, 
one finds
\eqna\IIIe
$$\eqalignno{ G_2&={1\over\alpha}-2{\partial\over\partial\alpha}F(\alpha,g)=
{1\over\alpha}(1+g G_4)& \IIIe a
\cr
G_2^{(1)}&={1\over\alpha}(1+g G_4^{(1)})& \IIIe b 
\cr }$$
which in particular proves that in the large $N$ limit $G_2=G^{(1)}_2$. 

In that ``planar'' limit, the function $F$ has been computed by a variety
of techniques: saddle point approximation \BIPZ, orthogonal polynomials \BIZ, 
``loop equations'' (see \DFGZJ\ for a review). 
With the current conventions and normalizations, 
\eqn\IIIf{
F(\alpha,g) = \oh \log a^2 -{1\over 24} (a^2-1)(9-a^2) } 
where $a^2$  is the solution of
%
\eqn\IIIg{3 {g\over\alpha^2} a^4 - a^2+1=0 } 
which is equal to $1$ for $g=0$. We have the expansion
\eqnn\IIIh 
$$\eqalignno{
F(1,g)
&=\oh \big(g+{9\over 4} g^2 +9 g^3 +{189\over 4}g^4
+\cdots\big)\cr
&= \sum_{p=1}^\infty (3g)^p{(2p-1)!\over p!(p+2)!}&\IIIh\cr }   $$
The 2- and 4-point functions are thus 
\eqna\IIIi
$$\eqalignno{
G_2&= G_2^\c={1\over 3\alpha} a^2(4-a^2)& \IIIi a 
\cr
G_4 &= {1\over\alpha^2}a^4(3-a^2) & \IIIi b 
\cr
G_4^\c&= G_4 -2 G_2^2 =-
{1\over 9\alpha^2} a^4 (a^2-1)(2a^2-5) & \IIIi c 
\cr}$$
where $G_4^\c$ is the connected 4-point function. 
More generally, $G_{2n}$ is of the form $\alpha^{-n}$ times a
polynomial in $a^2$.
The singularity of $a^2$ at $g/\alpha^2=1/12$ 
determines the radius of convergence 
of $F$ and of the $G_{2n}$.

\subsec{Removal of self-energies} 
\noindent
If we want to match Feynman diagrams contributing to the free energy 
$F^{(1)}$ with {\it prime} knots or links, we have to eliminate a 
certain number of redundancies. 
We have first to eliminate the ``self-energy insertions'' that 
correspond either to non prime knots or to 
irrelevant crossings (``nugatory'' in the language of Tait). 
This is simply achieved by choosing $\alpha$ as
a function of $g$ such that
\eqn\IIIj{G_2(\alpha(g),g)=1\ . }

The function $a^2(g):=a^2(\alpha(g),g)$ is obtained by eliminating $\alpha$
between Eqs. \IIIg\ and \IIIj
, i.e.
%
$$\eqalignno{
3 {g\over \alpha^2} a^4 - a^2+1 & =0  &\IIIg ' 
\cr
{1\over 3} a^2(4-a^2)&= \alpha  & \IIIj ' 
\cr }$$
which implies that $a^2(g)$ is the solution of
%
\eqn\IIIl{27 g= (a^2-1)(4-a^2)^2 }
equal to 1 when $g=0$; $\alpha(g)$ is then given by
%
\eqn\IIIm{ \alpha(g)={1\over 3} a^2(g) (4-a^2(g))\ . }
%
%
%
%
\fig{(a):  decomposition of the two-point function into
its one-particle-irreducible part; (b) discarding one-vertex-reducible
contributions}
{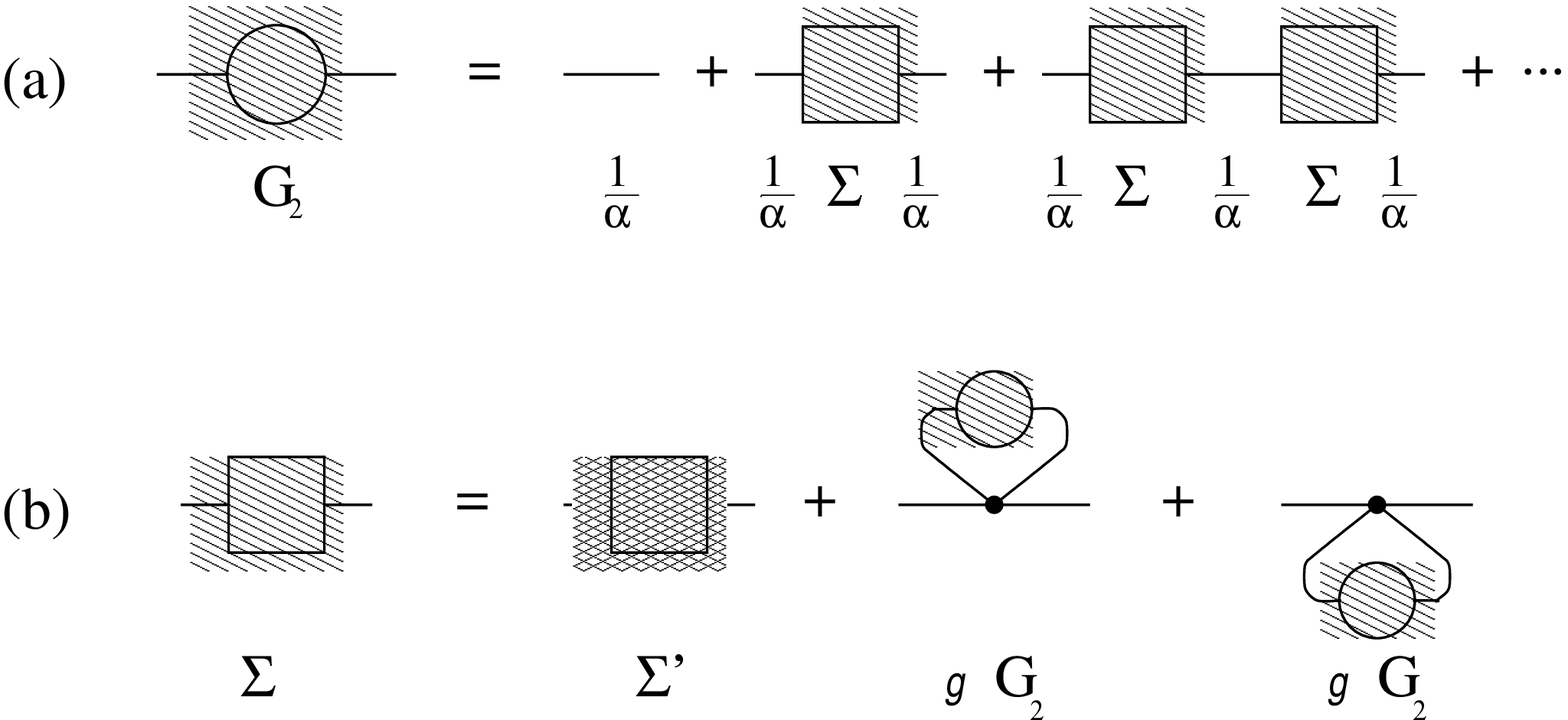}{8cm}\figlabel\twoptfn

Let us now  consider correlation functions. In field theory, 
it is common practice to define 
{\it truncated} diagrams, whose external lines carry no propagator, 
and {\it one-particle-irreducible } ones, that remain 
connected upon cutting of any line. 

The two-point function $G_2(\alpha,g)$
may be expressed in terms of the ``self-energy'' function $\Sigma$ 
\eqn\IIIn{
G_2(\alpha,g)={1\over\alpha-\Sigma(\alpha,g)}}
which is the sum of (non-trivial) truncated, one-particle-irreducible
graphs (Fig.~\twoptfn). One can further discard the contributions that are
one-vertex-reducible by defining $\Sigma'$ (see Fig.~\twoptfn(b))
%
\eqn\IIIo{\Sigma'(\alpha,g)=\Sigma(\alpha,g) -2 G_2(\alpha,g)}

If we now remove the self-energy insertions by imposing condition \IIIj, 
we find that Eqs. \IIIn\ and \IIIo\ 
simplify, so that
$\Sigma(g):=\Sigma(\alpha(g),g)$ and
$\Sigma'(g):=\Sigma'(\alpha(g),g)$ are given by
\eqna\IIIp
$$\eqalignno{
\Sigma(g)&=\alpha(g)-1& \IIIp a 
\cr
\Sigma'(g)&=\alpha(g)-1-2g& \IIIp b 
\cr
}$$
i.e. 
obtained from $\alpha(g)$ by removing the
first terms in its expansion in powers of $g$.

The procedure extends to all correlation 
functions.
Finally one obtains the corresponding ``free energy'' $F^{(1)}(g)$
by dividing the term of order $n$ of $\Sigma'(g)$ by $2n$
(= number of times of picking a propagator in a diagram
of the free energy to open it to a two-point function, cf Eq. 
\IIIe{} 
above).

We may also compute the function
$ \Gamma(\alpha,g) =G_2^\c(\alpha,g)^{-4} G_4^\c(\alpha(g),g)$, 
which counts the truncated (automatically one-particle-irreducible)
connected 4-point functions. After removal of the self-energy
insertions,
$\Gamma(g):=\Gamma(\alpha(g),g)$ becomes simply (Eqs. 
\IIIe{a}, \IIIi{c} and \IIIp{a}) 
\eqn\IIIq{\Gamma(g)={\Sigma'(g)\over g}=2{d\over dg} F^{(1)}(g)} 
Explicitly,
%
\eqn\IIIr{\Gamma(g)= -{1\over (4-a^2(g))^2}(a^2(g)-1)(2a^2(g)-5)}
%

Perturbatively one finds
\eqnn\IIIs
$$\eqalignno{
a^2(1,g)&=  1 +3g +18g^2 +135 g^3 +1134 g^4 +10206 g^5 +96228 g^6 
+\cdots \cr
G_2(1,g)&= 1 +2g +9 g^2 +54 g^3 +378 g^4 +2916 g^5 +24057 g^6 +\cdots
\cr
\alpha(g)&=1+2g+g^2 +2g^3 +6g^4 +22 g^5 +91 g^6 +\cdots & \IIIs \cr 
\Gamma(\alpha(g),g)&= g + 2 g^2  + 6 g^3  + 22 g^4  + 91 g^5  + 408 g^6 
 + \cdots \cr
%
F^{(1)}(g)&= {g^2\over 4}+{g^3\over 3}+{3 g^4\over 4 }
+{11 g^5\over 5} +{91 g^6\over 12 }+\cdots\cr }$$
%
\vskip -5pt %
%
\fig{(a): the first links, with the labelling of \Ro;  
(b) the corresponding diagrams
contributing to $F^{(1)}$. For simplicity, the diagrams are not
oriented, but the weights are those of the $(MM^\dagger)^2$ theory; 
(c) diagrams up to order 3 contributing to
$\Gamma$: the last four are pairwise flype-equivalent.}
{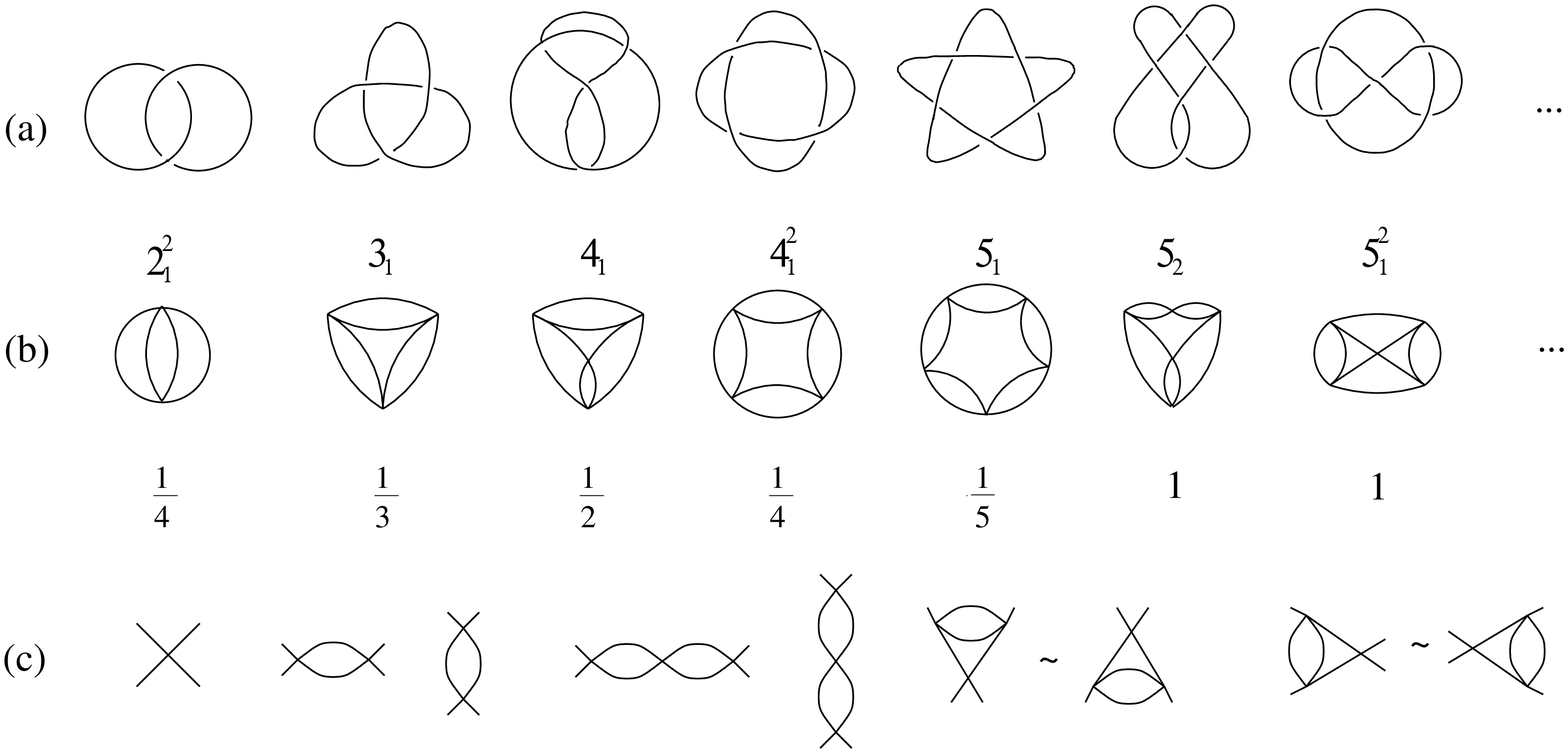}{12cm}\figlabel\links

The first terms of $F^{(1)}$ match 
the counting of the first prime links 
weighted by their symmetry factor (see Fig.~\links). 
%
\eqn\IIIs{ F^{(1)}(g)= {1\over 4} g^2 +{1\over 3} g^3
+({1\over 2}+{1\over 4})g^4+({1\over 5}+1+1)g^5 
+\cdots}
%
Starting with order $6$, however, there is an overcounting of links due
to neglecting the flype equivalence. 
The overcounting occurs already at order 3 if $\Gamma$ is used to 
count tangles.

\medskip
{\it Asymptotic behavior of the coefficients $f_n$  of 
$F^{(1)}(g)$}\par
\noindent
The singularity is now given by the closest zero 
of the equation $g/\alpha^2(g)=1/12$, which gives
$g_\star={4\over 27}$. Thus 
\eqn\IIIt{
f_n \sim {\rm const}\ b^n\, n^{-{7\over 2}}}
with 
%
\eqn\IIIu{ b={27\over 4}=6.75 \ . }
%


\subsec{Two-particle irreducibility}
\noindent Since the flype acts on tangles, we have to examine more closely 
the generating function $\Gamma(g)$ of
connected 4-point functions with no self-energies. We want to regard the
corresponding diagrams as resulting from the dressing of more 
fundamental objects. This follows a pattern familiar in  field theory, 
whose language we shall follow, while indicating in brackets the 
corresponding terminology 
of knot theory. 


We say that a 4-leg diagram (a tangle)
is {\it two-particle-irreducible} (2PI) (resp two-particle reducible, 2PR)
if cutting any two distinct propagators leaves it connected (resp
makes it disconnected). A 2PR diagram is thus the ``sum'' of smaller
components (see Fig.~\nprime). 
A {\it fully two-particle-irreducible} 
diagram is a 4-leg diagram such that any of its 4-leg subdiagrams
(including itself) is 2PI. Conversely a fully 2PR diagram (algebraic tangle)
is constructed by iterated sums starting from the 
simple vertex. 
We shall also make use of 
{\it skeletons}, which are  generalized diagrams (or ``templates'') 
in which some or all vertices are replaced by blobs (or ``slots'').
The concepts of fully 2PI skeleton (or basic polyedral template) 
and of fully 2PR skeleton (algebraic template) follow naturally, 
with however the extra condition that only blobs should appear in the former.

These blobs may then be substituted by 4-leg diagrams, resulting in a 
``dressing''. As will appear, the 
general 4-leg diagram (tangle)  results either from the dressing 
of a fully 2PI skeleton by generic 4-leg diagrams (type I tangles)
or  from the dressing of non trivial fully 2PR skeletons
 by  type I tangles. 
The action of flypes will be only on the  fully 2PR skeletons 
appearing in this iteration.

Because in the $(MM^\dagger)^2$ theory with 4-valent vertices,
there is no diagram which is two-particle-reducible in both channels, 
any diagram of $\Gamma(g)$ must be
\item{i)} 2-particle-irreducible in the vertical channel (V-2PI) but 
possibly 2-particle-reducible in the horizontal one. We denote
$V(g)$ the generating function of those diagrams.
\item{ii)} or 2-particle-irreducible in the horizontal channel (H-2PI) but 
possibly 2-particle-reducible in the vertical one. We denote
$H(g)$ the generating function of those diagrams.
\par\noindent 
Obviously there is some overlap between these two classes, corresponding 
to diagrams that are 2-particle-irreducible in both channels (2PI),
including the simple vertex. Let $D(g)$ denote their generating function. 
We thus have 
\eqn\IIIv{
\Gamma= H+V-D\ .}
Now since $V$ encompasses diagrams that are 2PI in both channels, 
plus diagrams that are once 2PR in the horizontal channel
(i.e. made of two H-2PI blobs joined by two propagators), 
plus diagrams twice H-2PR, etc, we have (see Fig.~5) 
\eqnn\IIIw
$$\eqalignno{ V&= D +H H + H  H  H +\cdots \cr
&= D+{H^2 \over 1-H }\ . & \IIIw 
\cr } $$
%
%
%
%
\fig{Graphical representation of Eqs. \IIIv\ and \IIIw.} 
{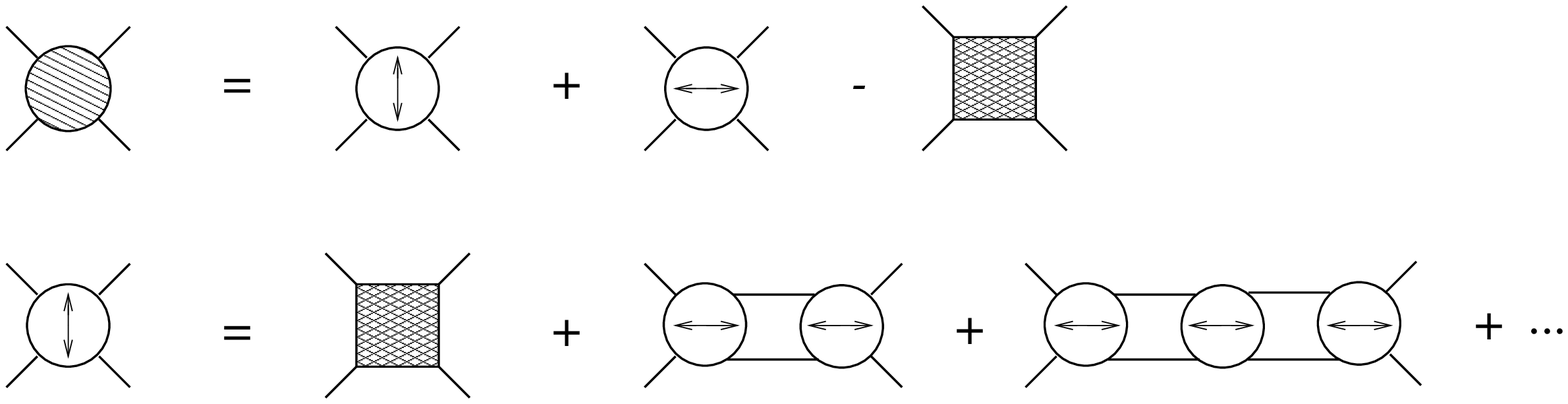}{8cm}\figlabel\skeleton
\noindent Since for obvious symmetry reasons, $H=V$, we have the pair 
of equations
\eqnn\IIIx
$$\eqalignno{ \Gamma&=2H-D\cr
H&= D+{H^2 \over 1-H }\ .&  \IIIx 
\cr }$$
Eliminating $H$ yields
\eqn\IIIy{
D =\Gamma{1-\Gamma \over 1+\Gamma }\ .} 
Eq. \IIIy\ 
can be inverted to allow to reconstruct the 
function $\Gamma(g)$ out of the 2PI function $D(g)$. For later
purposes, we want to distinguish between the single vertex diagram and
non-trivial diagrams:
\eqn\IIIz{
D(g)=g+\zeta(g)} 
In terms of these new variables, one has:
\eqn\IIIaa{
\Gamma(g)=\Gamma\{g,\zeta(g)\} }
(note the braces are used to distinguish the
different variables) where
\eqn\IIIab{
\Gamma\{g,\zeta\}= \oh\Big[
1-g-\zeta -\sqrt{(1-g-\zeta)^2-4(g+\zeta)} \Big] \ .}
is the generating function 
\eqn\IIIac{
\Gamma\{g,\zeta\}=\sum_{m,n} \gamma_{m,n} g^m \zeta^n} 
of the number of 
fully 2PR skeleton diagrams with $m$ vertices and $n$ blobs
(algebraic templates, including the trivial one made of a single blob 
and no vertex). Similarly,
one could define $H\{g,\zeta\}=V\{g,\zeta\}$, and of course
$D\{g,\zeta\}=g+\zeta$.
Note that the function $\Gamma\{g,\zeta\}$ does not depend on
the precise form of $\Gamma(g)$, since we have only used
the relation \IIIy\ 
which was derived from general diagrammatic
considerations.

Inversely, we shall need
the fully 2PI skeletons which are obtained from $\zeta(g)$ by replacing
subgraphs that have four legs with blobs.
Defining the inverse function $g[\Gamma]$ of $\Gamma(g)$,
the generating function of these skeleton diagrams is simply
$\zeta[\Gamma]:=\zeta(g[\Gamma])$, or more explicitly
\eqn\IIIad{
\zeta[\Gamma]=\Gamma{1-\Gamma\over 1+\Gamma} -g[\Gamma] 
\ .}

The function $g[\Gamma]$ satisfies by definition $\Gamma(g[\Gamma])=\Gamma$,
and is found
by eliminating $a^2$ between Eqs. $(11)$ and $(16)$. 
Setting $\eta=1-a^2$, we recover the system of \STh,~\Tutte:
\eqnn\IIIae
$$\eqalignno{
27 g&= -\eta (3+\eta)^2 & \IIIae \cr
\Gamma&= -{1\over (3+\eta)^2} \eta (3+2\eta)\cr }$$
which leads to
\eqn\IIIaf{
g[\Gamma]=\oh{1\over (\Gamma+2)^3}\Big[1+10\Gamma-2\Gamma^2
-(1-4\Gamma)^{{3\over 2}}\Big] }
and finally
\eqn\IIIag{
\zeta[\Gamma]=-{2\over 1+\Gamma} +2 -\Gamma -g(\Gamma)\ }
is the desired generating function of fully 2PI skeletons
(in the notations of \STh, this is $q(g)\,$).
The property mentionned above that $\Gamma$ is obtained by dressing
is expressed by the identity
\eqn\dress{\Gamma(g)=\zeta[\Gamma(g)] + \sum_{m,n\atop (m,n)\ne(0,1)}
\gamma_{m,n} g^m \zeta^n[\Gamma(g)] = \Gamma\{g,\zeta(g)\}\ .}
Perturbatively, we find 
\eqnn\IIIah
$$\eqalignno{
\Gamma(g)&=g+2g^2+6g^3+22g^4 + 91 g^5+\cdots \cr
D(g)&= g+g^5+10g^6 + 74 g^7+ 492 g^8 +\cdots \cr 
\Gamma\{g,\zeta\}&=g+\zeta + 2 (g+\zeta)^2
+ 6 (g+\zeta)^3  + 22 (g+\zeta)^4  + 90 (g+\zeta)^5+\cdots& \IIIah \cr
g[\Gamma]&= \Gamma-  2 \Gamma^2  + 2 \Gamma^3  - 2 \Gamma^4  + \Gamma^5  - 2 \Gamma^6  - 2 \Gamma^7  - 8 \Gamma^8  - 22 \Gamma^9  - 68 \Gamma^{10}   + 
\cdots\cr
\zeta[\Gamma]&= \Gamma^5+ 4 \Gamma^7
+6 \Gamma^8 + 24 \Gamma^9 + 66 \Gamma^{10}+\cdots\ .
\cr}$$
\vskip-10pt
%
%
\fig{The first contributions to $\zeta[\Gamma]$. All diagrams
can be obtained from the ones depicted by rotations of 90 degrees.  }
{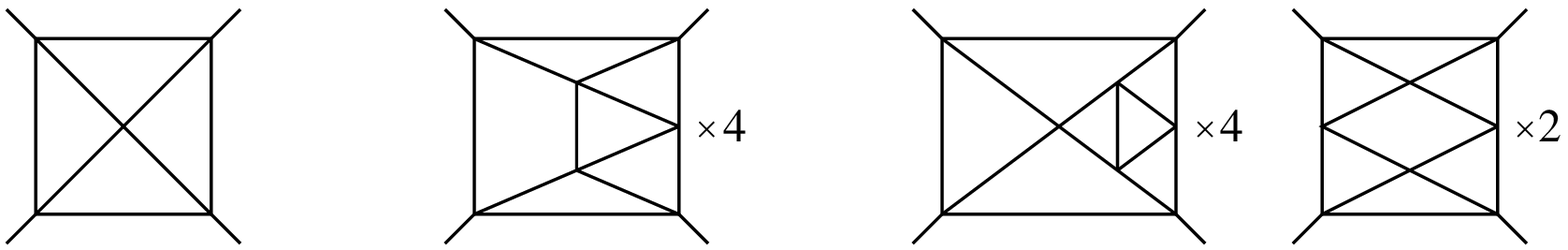}{10cm}\figlabel\tutte
\noindent The first contributions to $\zeta[\Gamma]$ are depicted 
on Fig.~\tutte.
The closest singularity of $g[\Gamma]$ and of $\zeta[\Gamma]$
is $\Gamma_\star=1/4$ which corresponds to
$g[\Gamma_\star]=g_\star=4/27$ and $\zeta[\Gamma_\star]=1/540$.

\subsec{Quotienting by the flype}  
\noindent The last 
step is to take into account the flype equivalence; it is
borrowed from the discussion of Sundberg and Thistlethwaite
and reproduced here for completeness.
The fully 2PR skeletons contained in Eq. \IIIab\ 
now have to be replaced with skeletons
in which the flype equivalence has been quotiented.
Then, they have to be dressed by 4-point 
2PI functions using Eq. \IIIag. 

Let  $\tilde{\Gamma}\{g,\zeta\}$ be the generating function
of the number $\tilde{\gamma}_{mn}$ of 
{\it flype-equivalence classes} of fully 2PR 
skeleton diagrams  (algebraic templates) with $m$
vertices 
 and $n$ blobs 
\eqn\IIIbb{
\tilde{\Gamma}\{g,\zeta\}=\sum_{m,n} \tilde{\gamma}_{m,n} g^m \zeta^n\ .}
Let $\tilde{H}\{g,\zeta\}$ (resp.\ $\tilde{V}\{g,\zeta\}$)
denote the generating function of 
flype-equivalence classes of 
skeletons which 
are 2PI in the horizontal (resp.\ vertical) channel, 
including the single blob and the single vertex.
In a way similar to the decomposition performed above for $\Gamma$,
cf Eq.~\IIIv, 
we write
\eqn\IIIai{
\tilde{\Gamma}=\tilde{H}+\tilde{V}-D} 
where $D\{g,\zeta\}=g+\zeta$.
The equation analogous to Eq.~\IIIw\ 
is
\eqnn\IIIaj
$$\eqalignno{
\tilde{V}&=D+g\tilde{\Gamma}+(\tilde{H}-g)^2+(\tilde{H}-g)^3+\cdots\cr
&=D + g \tilde{\Gamma}
+{(\tilde{H}-g)^2\over 1-(\tilde{H}-g)}& \IIIaj 
\cr}$$
where by flyping we can remove the simple vertices inside the
$\tilde{H}$ and put them as a single contribution $g\tilde{\Gamma}$.

As before, $\tilde{H}=\tilde{V}$ 
for symmetry reasons, and after eliminating it
one gets an algebraic equation for $\tilde{\Gamma}\{g,\zeta\}$
\eqn\IIIak{
\tilde{\Gamma}^2 -(1+g-\zeta)\tilde{\Gamma}
 +\zeta +g{1+g\over 1-g}=0}
with solution
\eqn\IIIal{
\tilde{\Gamma}\{g,\zeta\}
= \oh\left[ (1+g-\zeta) -\sqrt{(1-g+\zeta)^2-8 \zeta-8{g^2\over 1-g}}\right]
\ , }
which should be compared with Eq.~\IIIab. 

\medskip
The last step to get the generating function 
\eqn\IIIala{\tilde{\Gamma}(g):=\tilde{\Gamma}\{g,\tilde{\zeta}(g)\}}
of flype equivalence classes of (alternating, prime) tangles is to
define $\tilde{\zeta}(g)$ via
the relation \IIIag, 
i.e.
\eqn\IIIam{
\tilde{\zeta}(g)= \zeta[\tilde{\Gamma}(g)]}
(The reason we can use Eq. \IIIag\ 
without any modification is that fully
2PI skeleton diagrams are not affected by flypes.)
In other words, $\tilde{\Gamma}(g)$ is the solution to the implicit equation
\eqn\IIIan{
\tilde{\Gamma}(g)=\tilde{\Gamma}\{g,\zeta[\tilde{\Gamma}(g)]\}\ ,}
where $\zeta[\Gamma]$ and $\tilde{\Gamma}\{g,\zeta\}$ are
provided by Eqs.~\IIIaf, \IIIag\ and \IIIal, 
and which vanishes at $g=0$.

Perturbatively
\eqn\IIIao{
\tilde{\Gamma}(g)=g+2g^2+4g^3+10 g^4 +29 g^5+98 g^6+372 g^7+\cdots \
.}
Asymptotic behavior of $\tilde{\Gamma}(g)$: we know the closest
singularity of $\zeta[\Gamma]$;
we therefore set $\tilde{\Gamma}=\Gamma_\star=1/4$ and
$\tilde{\zeta}=\zeta_\star=1/540$ in Eq. \IIIak\ 
and solve for $g$:
\eqn\IIIap{
{\tilde g}_\star={-101+\sqrt{21001}\over 270}\ .}
This provides the asymptotic behavior of the coefficients $\tilde{f}_n$
of the ``free energy'' $\tilde{F}^{(1)}(g)$ defined by
$\tilde{\Gamma}(g)=2{d\over dg}\tilde{F}^{(1)}(g)$:
\eqna\IIIaq 
$$\eqalignno{
\tilde{f}_n &\sim {\rm const}\ \tilde{b}^n\, n^{-{7\over 2}}
&\IIIaq a \cr 
%
%
\tilde{b}&=(101+\sqrt{21001})/40\approx 6.147930\ . 
& \IIIaq b\cr }
$$ 
Note that while the first factor $\tilde{b}^n$ in the expansion
(``bulk free energy'') is non-universal, the second factor
$n^{-7/2}$ is universal (critical behavior of pure gravity \DFGZJ;
the exponent $\alpha=-{7\over2}$ gives the
``string susceptibility'' exponent $\gamma$ of non-critical
string theory \KM: $\gamma=\alpha+3=-\oh$)
and in particular is identical in Eqs.~\IIIt\ and \IIIaq{a}.
Note also that this free energy counts links with a weight 
that takes into account the equivalence under flypes and which is less easy 
to characterize. 

\newsec{Concluding remarks}
\noindent
In this paper we have reproduced the results of \STh\ using prior
knowledge of graph counting derived from matrix models,
while Sundberg and \Th\ were using results of Tutte \Tutte. Admittedly
the progress is modest. We hope, however, that our method
may give some clues on problems that are still open,
such as the counting of knots rather than links. To control 
the number of connected components of a link is in principle
easy in our approach. We should consider an integral over $n$
matrices $M_\alpha$, $\Ga=1,\cdots,n$ interacting through a term
$\sum_{\Ga,\Gb} (M_\Ga M_\Gb^\dagger)^2$
and look at the dependence of 
$F^{(1)}$ on $n$. The term linear in $n$ receives contributions
only from one-component diagrams, hence after a treatment similar 
to that of Sect.\ 3, it should give a generating function of
the number of knots, weighted as before by their symmetry factor.
Unfortunately, the computation of these matrix integrals
and their subsequent treatment (removal of self-energies
and flype equivalences) is for 
generic $n$ beyond our capabilities (see however
\KP\ for a first step in this direction).

Another problem on which matrix technology might prove useful would 
be in the counting of non alternating diagrams. But there the main 
problem is knot-theoretic rather than combinatorial: how 
does one get rid of multiple counting associated with Reidemeister moves?

\vskip1cm
\centerline{\bf Acknowledgements} 
It is a pleasure to acknowledge an informative exchange with V.~Jones
at the beginning of this work, and interesting discussions with 
M.~Bauer. 
Part of this work was performed when the authors were 
participating in the programme on Random Matrices and their Applications at
the MSRI, Berkeley. They want to thank Prof.\ D.~Eisenbud
for the hospitality of the Institute, and the organizers
of the semester, P.~Bleher and  A.~Its, for their invitation that
made this collaboration possible. 
P.Z.-J.\ is supported in part by the DOE grant DE-FG02-96ER40559.

\vskip1cm
\listrefs
\bye